\documentclass[english,aip,reprint]{revtex4-1}
\usepackage[LGR,T1]{fontenc}
\usepackage[latin9]{inputenc}
\usepackage{color}
\usepackage{babel}
\usepackage{graphicx}
\usepackage[unicode=true,
 bookmarks=true,bookmarksnumbered=false,bookmarksopen=false,
 breaklinks=true,pdfborder={0 0 1},backref=false,colorlinks=true]
 {hyperref}
\hypersetup{pdftitle={Cryogenic preamplification of a single-electron-transistor using a silicon-germanium heterojunction-bipolar-transistor},
 pdfauthor={M. J. Curry},
 allcolors=blue}

\makeatletter

\DeclareRobustCommand{\greektext}{%
  \fontencoding{LGR}\selectfont\def\encodingdefault{LGR}}
\DeclareRobustCommand{\textgreek}[1]{\leavevmode{\greektext #1}}
\DeclareFontEncoding{LGR}{}{}
\DeclareTextSymbol{\~}{LGR}{126}

\@ifundefined{textcolor}{}
{%
 \definecolor{BLACK}{gray}{0}
 \definecolor{WHITE}{gray}{1}
 \definecolor{RED}{rgb}{1,0,0}
 \definecolor{GREEN}{rgb}{0,1,0}
 \definecolor{BLUE}{rgb}{0,0,1}
 \definecolor{CYAN}{cmyk}{1,0,0,0}
 \definecolor{MAGENTA}{cmyk}{0,1,0,0}
 \definecolor{YELLOW}{cmyk}{0,0,1,0}
}

\makeatother

\begin{document}

\title{Cryogenic preamplification of a single-electron-transistor using
a silicon-germanium heterojunction-bipolar-transistor}

\author{M. J. Curry}

\affiliation{Department of Physics and Astronomy, University of New Mexico, Albuquerque,
New Mexico, 87131, USA}

\affiliation{Center for Quantum Information and Control, University of New Mexico,
Albuquerque, New Mexico, 87131, USA}

\affiliation{Sandia National Laboratories, 1515 Eubank Blvd SE, Albuquerque, New
Mexico, 87123, USA}

\author{T. D. England}

\author{N. C. Bishop}

\author{G. Ten-Eyck}

\author{J. R. Wendt}

\author{T. Pluym}

\author{M. P. Lilly}

\affiliation{Sandia National Laboratories, 1515 Eubank Blvd SE, Albuquerque, New
Mexico, 87123, USA}

\author{S. M. Carr}

\affiliation{Center for Quantum Information and Control, University of New Mexico,
Albuquerque, New Mexico, 87131, USA}

\affiliation{Sandia National Laboratories, 1515 Eubank Blvd SE, Albuquerque, New
Mexico, 87123, USA}

\author{M. S. Carroll}

\affiliation{Sandia National Laboratories, 1515 Eubank Blvd SE, Albuquerque, New
Mexico, 87123, USA}
\begin{abstract}
We examine a silicon-germanium heterojunction bipolar transistor (HBT)
for cryogenic pre-amplification of a single electron transistor (SET).
The SET current modulates the base current of the HBT directly. The
HBT-SET circuit is immersed in liquid helium, and its frequency response
from low frequency to several MHz is measured. The current gain and
the noise spectrum with the HBT result in a signal-to-noise-ratio
(SNR) that is a factor of $10-100$ larger than without the HBT at
lower frequencies. The transition frequency defined by SNR = 1 has
been extended by as much as a factor of 10 compared to without the
HBT amplification. The power dissipated by the HBT cryogenic pre-amplifier
is approximately 5 nW to 5 \textgreek{m}W for the investigated range
of operation. The circuit is also operated in a single electron charge
read-out configuration in the time-domain as a proof-of-principle
demonstration of the amplification approach for single spin read-out.
\end{abstract}
\maketitle
Donor spin qubits have recently received increased interest because
of the demonstration of high fidelity coherent control of phosphorus
donors using a local electron spin resonance technique.\cite{Pla2012,Witzel2010}
This approach is of interest both for quantum information\cite{Awschalom2013,Zwanenburg2013,Morton2011}
as well as representing a new experimental platform to investigate
the behavior of single impurities in semiconductors using electron
and nuclear magnetic resonance. Single-shot readout\cite{Elzerman2004,Hanson2005,Barthel2009}
of the spin polarization is an important component of the measurement.
It may be accomplished using a wide-band measurement of the single
electron transistor\cite{Devoret2000} (SET) conductance, which is
sensitive to the ionization condition of any nearby donors.\cite{Morello2010,Tracy2013}
The technique relies on alignment of the neighboring SET chemical
potential between discrete Zeeman energy levels. The donor spin-up
electron ionizes into the SET, leading to a detectable transient change
in the local electrostatic potential, while a SET electron waits to
reload into the donor as a spin-down. The temporary ionization of
the donor changes the conductance of the SET, which is measured as
a current pulse corresponding to a spin-up electron or no pulse if
the electron was spin-down.

Read-out fidelity can be no better than what the signal-to-noise-ratio
(SNR) provides for a particular bandwidth, although other factors
can introduce errors that degrade the fidelity, such as rapid tunneling
events that are faster than the bandwidth of the read-out. The donor
read-out technique is performed at cryogenic temperatures less than
4 K, which is typically necessary to observe the spin read-out of
the donor state at reasonably low magnetic fields. The SET current
is subsequently amplified at room-temperature (RT) using one or several
amplification stages, typically including a transconductance amplifier.
The line capacitance between the transconductance amplifier and the
SET typically sets the limits of performance of the circuit. Increased
read-out bandwidth can improve fidelity, for example, by detecting
faster tunnel events, however, the increased bandwidth reduces SNR.
The SNR can be increased if amplification is introduced before the
dominant noise source contributes to the signal.

Several approaches have been pursued to maximize SNR using cryogenic
electronics for read-out and amplification. One technique is to embed
an SET in a RF resonant circuit, referred to as RF-SET,\cite{Schoelkopf1998,Barthel2010,Villis2014}
which has resulted in some of the most competitive read-out performance.
However, the RF-SET technique requires a significant investment to
implement, it introduces some challenges to integration,\cite{Gonzalez-Zalba2015,Verduijn2014,Colless2013}
and for the purpose of donor spin read-out it can introduce an additional
complication of directly modulating the chemical potential of the
SET. An alternative technique, similar in some respects to the RF-SET
approach, is to couple a SET or similar device to a superconducting
resonator,\cite{Wallraff2004,Petersson2012,Schmidt2014} which may
be followed by additional superconducting quantum circuitry such as
a Josephson Parametric Amplifier.\cite{Vijay2009,Abdo2014,Stehlik2015}
Current comparators have shown promise but their thresholds of sensitivity
have been near the limits of the current output of SETs, making them
difficult to implement without a preamplification stage.\cite{Gurrieri2008,Das2014}
Cryogenic preamplification using discrete high-electron-mobility-transistors
(HEMTs) has been investigated resulting in sufficient SNR for a particular
bandwidth.\cite{Vink2007} However, HEMTs may require a relatively
high power dissipation, and the typical circuit configuration introduces
a fixed load resistance in front of the gate that can limit the circuit
bandwidth.

In this letter, we use a discrete, commercial silicon-germanium (SiGe)
heterojunction bipolar transistor\cite{Cressler2002,Cressler2012,Cressler2010,Najafizadeh2009,Joseph1995}
(HBT) for cryogenic\cite{Balestra2001,Gutierrez2000} amplification
of a silicon SET\textquoteright s output current. This is a preamplification
stage for a single electron spin read-out circuit. The SiGe HBT can
be operated at relatively low power, has a low overhead for implementation,
and, in principle, could be integrated with a silicon-based qubit process
flow. We find that the HBT provides a current gain of order $100-2000$.
The current gain and the noise spectrum with the HBT result in a SNR
that is a factor of $10-100$ larger than without the HBT at lower
frequencies. The transition frequency defined by SNR $=1$ has been
extended by as much as a factor of 10 compared to without the HBT
amplification. The power dissipated by the HBT is estimated to be
between 5 nW and 5 \textgreek{m}W for the relevant operation range.

\begin{figure}
\includegraphics[width=8.5cm]{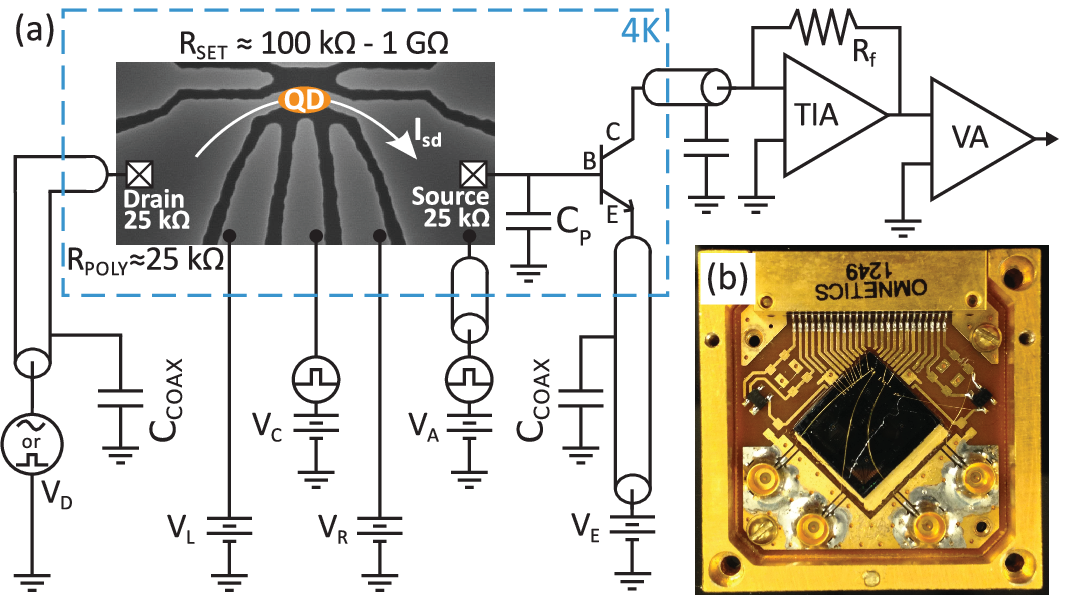}

\protect\caption{The HBT-SET circuit. The SEM image shows the silicon metal-oxide-semiconductor
device geometry with polysilicon gates labeled $V_{L}$,$V_{C}$,$V_{R}$,
and $V_{A}$. The quantum dot (QD) is formed beneath the narrow channel
of the gate labeled $V_{A}$. The circuit is DC biased by $V_{E}$
and AC biased by $V_{D}$ (either sinusoidal or pulsing inputs). The
parasitic capacitance, $C_{P}$, is due to the device and/or PCB.
The parasitic capacitance, $C_{COAX}$, is due to the length of the
wires leading to and from the device immersed in liquid helium at
4 K. Room-temperature transimpedance (TIA) and voltage (VA) amplifiers
are used to amplify the signal before it is read out on a lock-in
amplifier or oscilloscope.\label{fig:HBT-SET-circuit}}
\end{figure}

The measurement circuit with both HBT and SET is shown in Figure \ref{fig:HBT-SET-circuit}.
The base of the HBT is connected to the source of the SET using a
bond wire between a surface mount HBT and the SET chip, both of which
are immersed in liquid helium during measurement. The HBT collector
is connected to a one meter long Lakeshore 304 stainless steel braided
coaxial cable with a capacitance of approximately 174 pF/m. The
emitter is connected via an identical cable to either a Keithley 2400
or an Agilent 33500B for low and high frequency measurements, respectively.
The gate labeled $V_{A}$ is also connected to coaxial cables for
pulsed measurements, while all other leads were connected through twisted
pair lines. The HBT collector is connected to a room-temperature Femto
DLPCA-200 transimpedance preamplifier unless otherwise noted. A subsequent
SR560 voltage preamplifier is used as a variable bandwidth filter
but otherwise is set to a gain of 1. Lock-in measurements were done
with a Zurich HF2LI or SR830 using a 100:1 resistive voltage divider
and typically an excitation voltage of 100 \textgreek{m}V or 1 mV
on the SET drain, without or with the HBT, respectively, unless otherwise
noted. The DC bias was set by the DC source applied to the HBT emitter
with no voltage division.

\begin{figure}
\includegraphics[width=8.5cm]{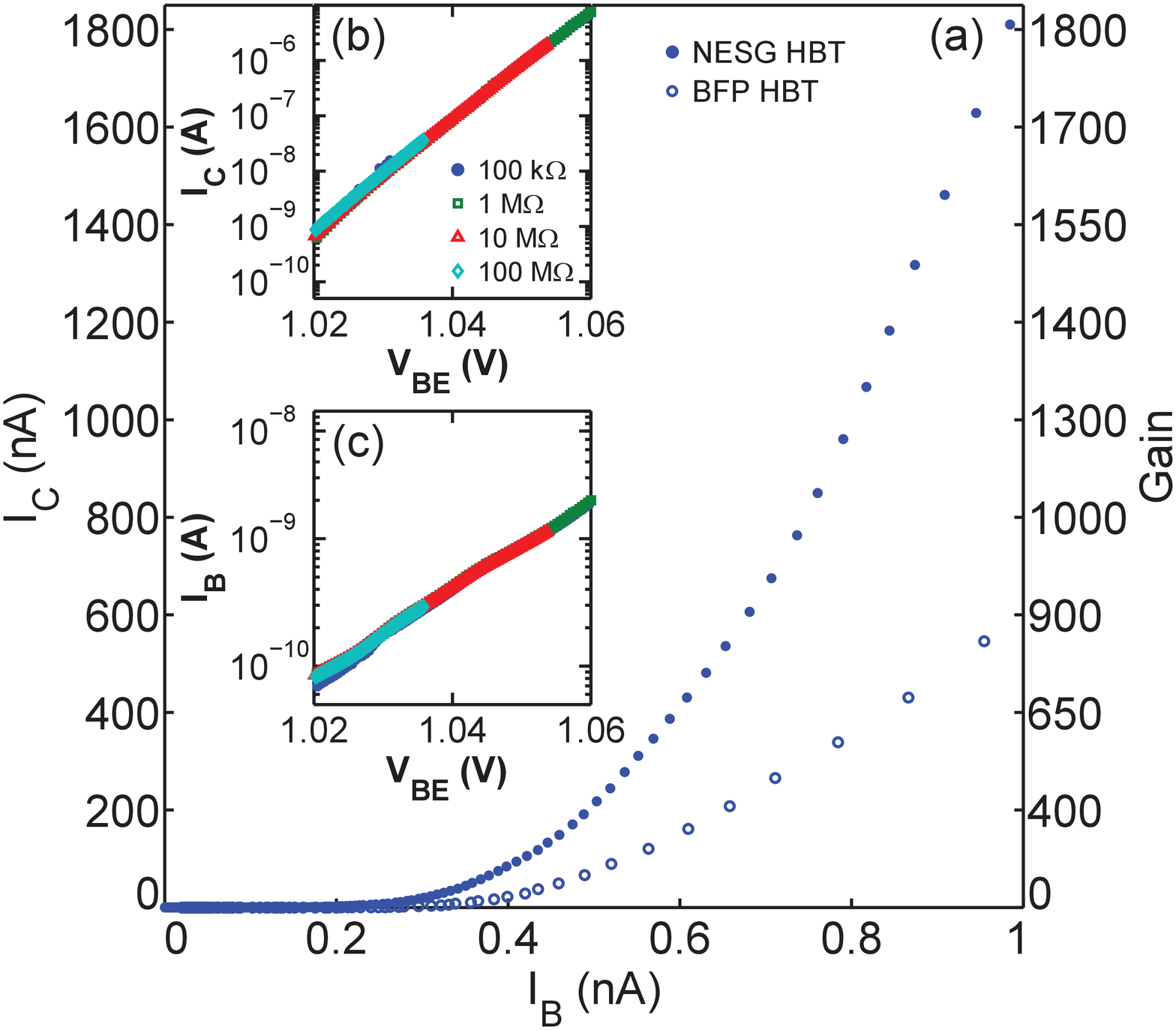}

\protect\caption{HBT biasing calibration curves at 4 K. (a) Collector current as
a function of base current for the CEL NESG3031M05 HBT used and an
Infineon BFP842ESD HBT. This curve enables mapping from read-out current
to device input current regardless of SET resistance. The gain shown
is for the NESG HBT. (b) Collector current as a function of the HBT
base-emitter voltage for different resistors in-line with an NESG
HBT. Identical HBT turn-on behavior is observed regardless of the
load resistance before the base-emitter junction of the HBT. (c) Base
current as a function of the HBT base-emitter voltage for different
resistors in-line with an NESG HBT. These current curves similarly
overlap.\label{fig:HBT-biasing}}
\end{figure}

HBTs were first characterized in liquid helium with room-temperature
load resistors without the SET to simulate different SET resistances
and calibrate the transistor\textquoteright s collector current as
a function of base current and load resistance, Figure \ref{fig:HBT-biasing}.
Multiple commercially available high bandwidth HBTs were measured
at low temperature. Resistances between 100 k\textgreek{W} and 1 G\textgreek{W}
were examined. The DC behavior of these HBTs at low temperature is
exponential. As the input current, $I_{B}$, increases, the readout
current, $I_{C}$, increases exponentially, Figure \ref{fig:HBT-biasing}(a).
The turn-on behavior of the HBT in our circuit depends solely on the
forward-bias diode drop across the base-emitter (BE) junction, $V_{BE}$.
$V_{BE}$ is calculated by subtracting the voltage drop across the
resistor from the bias applied to the emitter. For different resistances,
input and readout current behave exactly the same as $V_{BE}$ is
increased, Figure \ref{fig:HBT-biasing}(b) and \ref{fig:HBT-biasing}(c). Therefore, for a
given readout current, the input current is known and by using this
curve as a calibration, the potential across a SET connected to the
HBT, for a fixed emitter bias, can be estimated as $\triangle V_{SET}(I_{C})=|V_{E}|-|V_{BE}(I_{C})|$.
We note that not all HBTs measured at 4 K showed greater than unity
current gain ($I_{C}/I_{B}>1$) combined with correspondingly low
voltage of $0.1-2$ mV across the test resistance. Typical operation
of the silicon SETs for read-out is done with a bias voltage of $80-300$
\textgreek{m}V, well below the charging energy of the SET to avoid
reduction of sensitivity from broadening of the Coulomb blockade peaks.
Out of 25 HBTs characterized at 4 K, the California Eastern Laboratories
(CEL) NESG3031M05%
\footnote{At the time of publication, the manufacturer's website states that
the CEL NESG3031M05 HBT is no longer in production. However, the Infineon
BFP842ESD HBT had a similar biasing calibration curve at 4 K as shown
in Figure \ref{fig:HBT-biasing}(a).%
} HBT showed the highest current gain and lowest test resistance biasing,
so it was selected for measurements with the SET.

\begin{figure}
\includegraphics[width=8.5cm]{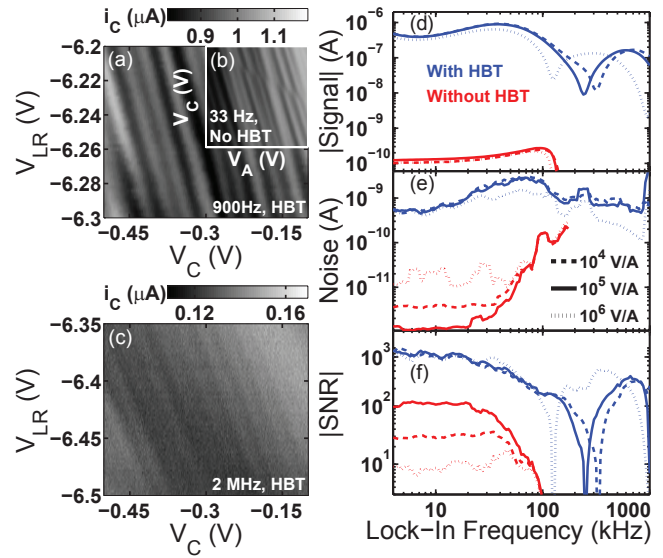}

\protect\caption{Stability plots and narrow band frequency domain data. All measurements
were performed at a temperature of 4 K. (a) Stability plot showing
the SET Coulomb blockade behavior of the HBT-SET circuit at 900 Hz.
Well defined peaks and contrast are found when an HBT is added. (b)
Stability plot of the same SET and resulting Coulomb blockade with no
HBT. Similar contrast to (a) is observed. (c) Stability plot showing
the SET Coulomb blockade in a very similar voltage range as (a) but
with an input frequency of 2 MHz. The SNR decreases at higher
frequencies in this narrow band measurement. (d)-(f) Narrow band
measurements as a function of input frequency. All data are the lock-in
output\textquoteright s in-phase quadrature. The signal and SNR are
plotted as absolute values.\label{fig:narrow-band}}
\end{figure}

To examine the frequency response of the SET with and without the
HBT, narrow band lock-in measurements were done by inputting a small
voltage sinusoidal signal into the SET\textquoteright s drain resulting
in a sinusoidal input current, $i_{b}$, and a sinusoidal readout
current, $i_{c}$. To ensure the DC operating point was minimally
perturbed, input signal magnitudes were constrained to $i_{c}\leq0.2\cdot I_{C}$,
remaining within a linear signal regime. A set of charge stability
plots show Coulomb blockade through the quantum dot (QD), Figures \ref{fig:narrow-band}(a)-\ref{fig:narrow-band}(c).
The stability plots are formed by sweeping the center plunger, $V_{C}$,
and stepping the left and right plungers, $V_{L,R}$, as indicated
in Figure \ref{fig:HBT-SET-circuit}. Qualitatively, the presence
of Coulomb blockade confirms that a DC bias can be chosen that produces
$V_{SD}$ sufficiently below the charging energy of the QD. We estimate
that $V_{SD}$ for $V_{E}=-1.051$ V is approximately 1 mV, extracted
from the appropriate $I_{C}$ vs $V_{BE}$ curve.

The SET-HBT current does not go to zero in the blockaded regions.
Verilog-A simulations of the circuit including a model for SET conductance
and the calibrated 4 K HBT parameters\cite{Mahapatra2004} predict
that the HBT-SET minimum conductance for the Coulomb blockade will
be prevented from going to zero. This behavior is believed to be a
consequence of having a floating source that increases $V_{SD}$ to
maintain some current through both the HBT and SET at all times. That
is, relatively small changes in the DC bias current through the HBT
lead to relatively large voltage shifts from $V_{BE}$ to $V_{SD}$.
The current turn-off of the Coulomb blockade is consequently suppressed
because the shift of voltage drop from $V_{BE}$ to $V_{SD}$ is always
enough to maintain a small current through both the SET and the HBT.
Quantum point contacts (QPC) are also frequently used as charge sensors
and would likely minimize this complication of the amplification circuit.
That is, the conductance of a QPC varies much less over similar bias
ranges, and usually very low conductance conditions can be avoided.

Quantitatively, the cases with and without the HBT are compared for equal
input signal, 100 \textgreek{m}V, and similar SET resistance, 100
k\textgreek{W} to 1 M\textgreek{W}, Figures \ref{fig:narrow-band}(d)-\ref{fig:narrow-band}(f).
The lock-in signal shown is the in-phase quadrature. The frequency
dependence of the narrow band SNR is shown for several room-temperature
transimpedance amplifier gain settings, Figure \ref{fig:narrow-band}(f).
The current gain and the noise spectrum with the HBT result in a SNR that is a factor of $10-100$ larger than without the HBT at
lower frequencies. For stability plots such as Figure \ref{fig:narrow-band}(a),
we were able to reduce the lock-in time constant by at least a factor
of 10 due to the increased SNR, thereby reducing the total acquisition
time by at least the same factor.

In Figure \ref{fig:narrow-band}(f), the transition frequency, defined
by SNR $=1$, for the case without the HBT is observed to be $\sim100$
kHz, where |SNR|$<<1$ for frequencies higher than 100 kHz. The transition
frequency is extended with the addition of the HBT. The extended transition
frequency with the HBT enabled acquisition of narrow band stability
plots at 2 MHz, as shown in Figure \ref{fig:narrow-band}(c). For
frequencies near and beyond approximately 200 kHz in Figure \ref{fig:narrow-band}(f),
a significant shift in phase that approaches 180\textdegree{} is observed
in the lock-in detected signal. The absolute value of the signal and
SNR is plotted in Figure \ref{fig:narrow-band}(d) and \ref{fig:narrow-band}(f)
in order to show this phase shift. Circuit analysis of Figure \ref{fig:HBT-SET-circuit}(a)
indicates that the HBT circuit has a pole due to the SET resistance
and the parasitic capacitance, $C_{P}$, as well as an additional
pole due to the HBT base-collector resistance and the parasitic capacitance
of the coaxial cables, $C_{COAX}$.

\begin{figure}
\includegraphics[width=8.5cm]{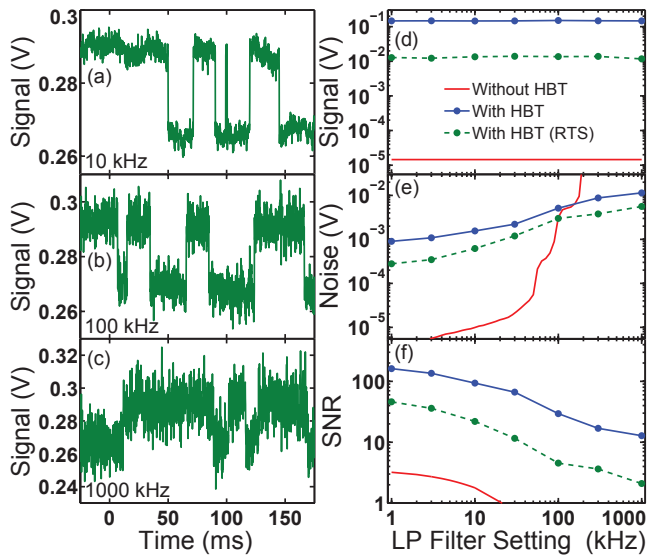}

\protect\caption{(a)-(c) Single-shot oscilloscope traces of RTS due to proximal charge center tunneling. Room-temperature low
pass filter settings of 10 kHz, 100 kHz, and 1000 kHz are used to
monitor the RTS. (d)-(f) Measured wide band signal magnitude,
noise magnitude, and SNR as a function of low-pass (LP) filter setting.
The blue curves show calibrated input current (150 pA) pulse data
for the HBT-SET circuit. The red curve is the same current input as
the blue curve with noise calculated from the narrow band noise spectral
density for the SET circuit without the HBT. The green curve shows
the measured RTS data from Figures \ref{fig:wide-band}(a)-\ref{fig:wide-band}(c).\label{fig:wide-band}}
\end{figure}

The response of the SET-HBT circuit was measured in the time domain
by tuning the SET in resonance with a nearby charge center such that
tunneling events on/off of the charge center are observed as changes
in the conductance of the SET between two conductance states: charge
center neutral or ionized, Figures \ref{fig:wide-band}(a)-\ref{fig:wide-band}(c).
This is similar to a charge sensing or spin read-out configuration.
The magnitude of conductance change is not the largest possible but
was chosen in this case because of a combination of factors including
the average tunneling rate of the transition. These data were acquired
using a room-temperature amplification chain consisting of a Femto
DLPCA-200 transimpedance amplifier with sensitivity $10^{5}$ V/A
followed by an SR 560 voltage amplifier with gain of 1 and variable
low-pass filter 3-dB frequencies. Measurements of signal amplitude,
noise (RMS deviation from the mean value of a voltage level), and
SNR are summarized for voltage-amplifier low-pass filter settings
up to 1 MHz. Other circuit parasitics introduce signal loss at frequencies
less than 1 MHz as indicated by the narrowband measurements, however,
SNR $>1$ is still achieved at $\sim1$ MHz with the HBT-SET circuit,
as shown by the green random telegraph signal (RTS) curve in Figure \ref{fig:wide-band}(f).

The circuit response to direct pulsing on the drain ohmic was investigated
to more directly examine rise/fall time response. The pulses were
generated by applying externally controllable square voltage pulses
to the drain ohmic, with the voltage amplitude converted to a current
amplitude by measuring and taking the difference of the current at
both voltage levels with a current meter. Square pulses with amplitude
150 pA and width 2 ms were used for the uppermost (blue) curve in
Figure \ref{fig:wide-band}(d). These data were acquired using a RT
amplification chain consisting of a Femto DLPCA-200 transimpedance
amplifier with sensitivity $10^{5}$ V/A followed by an SR 560 voltage
amplifier with a gain of 1 and variable low-pass filter 3-dB frequencies.
We found that the conductance of the SET modified the bandwidth of
the circuit response and that for the maximum and minimum conductances
examined, 0.313 \textgreek{m}S \& 0.183 \textgreek{m}S, respectively,
we found response times of 14.5 \textgreek{m}s and 0.612 \textgreek{m}s,
respectively. The SNR from the RTS like
behavior is overlaid as green curves. The blue curve in Figure \ref{fig:wide-band}(f)
shows an improved SNR primarily because the current through the SET
is being driven by an external pulse instead of being limited by the
change in conductance in the SET from the RTS charge center at a fixed
$V_{SD}$ range. Through the calibration of voltage to current, we
find a current amplitude of 13 pA for the RTS data in Figures \ref{fig:wide-band}(a)-\ref{fig:wide-band}(c).

For comparison between the device with and without the HBT, we estimate
the wide band SNR from the narrow band measurements. That is, we
assume the same current pulse amplitude of 150 pA is used from the
uppermost (blue) curve in Figure \ref{fig:wide-band}(d), and we calculate
the total noise by integrating the noise spectral density that was
measured in the narrow band measurements without the HBT. Noise spectral
density was calculated by dividing the solid red curve in Figure \ref{fig:narrow-band}(e)
by the square root of the noise-equivalent-power bandwidth of 1.25
Hz, calculated from the lock-in time constant and filter roll-off.
The noise spectral density was integrated to an upper limit equal
to the voltage amplifier low-pass filter 3-dB frequency, resulting
in the red curve in Figure \ref{fig:wide-band}(e). We find that the
noise increases nonlinearly with the HBT and also for the calculated
case without the HBT, while the signal stays constant throughout the
low-pass filter setting range. With the HBT, there is an increase
in SNR of about a factor of 10 for the RTS case (green), and about
a factor of 60 for the direct pulsing case (blue) at lower frequencies.
The greater SNR, particularly at lower frequencies, appears to be
due to an amplification in signal before the dominant noise source
is introduced, perhaps near the input of the preamplifier at room-temperature.
The SNR for the direct pulsing (blue) and the RTS (green) is reduced
at higher filter settings because of the nonlinearly increasing noise.
However, with the HBT, the gain is sufficiently high such that pulsing
events are detectable at the microsecond time scale.

An estimate of the DC power dissipation of the HBT can be made by
taking the product of the current through and voltage across the HBT.
The peak conductance conditions observed in this work correspond to
$I_{B}$ \ensuremath{\approx} 1 nA and $I_{C}$ \ensuremath{\approx}
5 \textgreek{m}A maximum and |$V_{E}$| \ensuremath{\approx} 1 V,
from which we can estimate a power dissipation of $V\cdot I=1(V)\times5\cdot10^{-6}(A)=5$
\textgreek{m}W. Regions off of peak conductance, that is, most of
the stability diagram, correspond to power dissipations as low as
10-100 nW. Even at the highest estimated power of 5 \textgreek{m}W,
the power dissipation is less than or about equal to the cooling power
of the lowest temperature stage of a dilution refrigerator.

We examined a discrete, commercial SiGe HBT for low power cryogenic
preamplification of a SET charge sensing circuit. The HBT-SET charge
sensing circuit is shown to produce a substantial increase in SNR
relative to the SET charge sensing circuit without an HBT. The gain
is non-linear when using the SET read-out configuration. Read-out
behavior is simulated by using the circuit to detect random telegraph
signal of a nearby charge center. The HBT-SET circuit was voltage
biased to a point where the power dissipated was $0.01-5$ \textgreek{m}W;
the gain was $100-2000$; and the source-drain bias across SET was
$\sim0.1-1$ mV. The current gain and the noise spectrum with the
HBT result in a SNR that is a factor of $10-100$ larger than without
the HBT at lower frequencies. The transition frequency defined by
SNR = 1 has been extended by as much as a factor of 10 compared to
without the HBT amplification. The increased performance is believed
to be due to signal gain near the SET before a major noise source
is introduced in front of the room-temperature transimpedance amplification
stage.

\bibliographystyle{aipnum4-1}
\bibliography{revised_manuscript.bbl}

\end{document}